\documentclass[12pt]{article}
\begin{document}

\pagestyle{empty}
\begin{flushright}
UNIL-IPT-03-1\\ 
March 2003
\end{flushright}
\vspace*{5mm}

\begin{center}

{\Large\bf QED from six-dimensional vortex and gauge anomalies}

\vspace{1cm}

{\large Seif Randjbar-Daemi$^a$\footnote{Email: seif@ictp.trieste.it}
and Mikhail Shaposhnikov$^b$\footnote{Email:
mikhail.shaposhnikov@ipt.unil.ch}}\\

\vspace{.6cm}

{\it{$^a$ The Abdus Salam ICTP, Strada Costiera 11, 34014 Trieste,
Italy}}

{\it {$^b$ Institute of Theoretical Physics, University of Lausanne,\\ 
CH-1015 Lausanne, Switzerland}}
\vspace{.4cm}
\end{center}

\vspace{1cm}  

\begin{abstract}

Starting from an anomaly-free Abelian Higgs model coupled to gravity
in a 6-dimensional space-time we construct an effective
four-dimensional theory of charged fermions interacting with U(1)
Abelian gauge field and gravity, both localised near the core of a
Nielsen-Olesen vortex configuration. We show that an anomaly free
theory in 6-dimensions can give rise to an anomalous  theory in D=4,
which suggests a possibility of consistent regularisation of abelian
anomalous chiral gauge theories in four dimensions. We also show that
the spectrum of charged bulk fermions has a mass gap. 

\end{abstract}

\vfill

\eject
\pagestyle{empty}
\setcounter{page}{1}
\setcounter{footnote}{0}
\pagestyle{plain}

\section{Introduction}

The idea that our observable 4-dimensional universe may be a brane
extended in some higher-dimensional space-time has a long history 
\cite{Rubakov:bb}-\cite{Randall:1999vf} and has been the subject of
many recent studies (for a review see, e.g. \cite{Rubakov:2001kp}). 
To implement this idea in practical terms one needs to find physical
mechanisms which localise the higher-dimensional fields to a
3-dimensional brane whose world volume is our 4-dimensional
space-time. D-branes of string theory \cite{Polchinski:1995mt} may
provide a natural mechanism for localisation of the fields of the
standard model. It is natural to ask whether local quantum field
theory formulated in higher dimensional space-time can achieve this
goal.

It was noticed already in~\cite{Rubakov:bb} that this is not an easy
task. Although the localisation of fermionic fields is quite easy to
achieve, it is much more difficult to obtain localised massless gauge
fields minimally coupled to the fermions living on a brane. A purely
field-theoretical mechanism for localisation of gauge fields based on
confinement was proposed in \cite{Dvali:1996xe}; it is still not
clear if it can be realised, as the higher-dimensional gauge theories
leading to confinement in the bulk and its absence on a brane are not
known (it may work, however, in a non-field theoretical  model of
confinement, \cite{Dubovsky:2001pe}).  Gravitational interactions in
higher dimensions help to improve the situation. For example, a local
string-like defect in six dimensions \cite{Gherghetta:2000qi}
provides localisation of the gauge field that has only gravitational
interactions with the string \cite{Oda:2000zc}. A six-dimensional
model in which the sixth coordinate is non-compact {\em \`a la}
Randall-Sundrum \cite{Randall:1999vf} and the fifth is compact {\em
\`a la} Kaluza-Klein does the same job \cite{Dubovsky:2000av}. 

In fact, the existence of a localised zero mode for the gauge field
is not enough for construction of a lower dimensional effective
theory - it is required that the modes living in the bulk interact
weakly with the mode living on a brane. It appears that the spectrum
of the bulk gauge modes localised by gravity is gapless
\cite{Dubovsky:2000av}, and, therefore, the viability of 
four-dimensional effective theory is questionable. The arguments that
it can be valid for the Abelian case were presented in
\cite{Dubovsky:2001pe}, whereas what happens in non-Abelian case is
obscure. 

The gauge fields discussed in \cite{Oda:2000zc,Dubovsky:2000av} were
external to the fields forming a topological defect. In the case of
an infinitely thin string, the metric solution has an SO(2) isometry
group and the corresponding metric component $h_{\mu\theta}$, with
$\theta$ being an angular coordinate, can play the role of the U(1)
gauge field \cite{Neronov:2001br}. In fact, a natural field theory
realisation of a string in six-dimensional space-time is the
Nielsen-Olesen  vortex \cite{Nielsen:qs}. This solution admits
gravity localisation \cite{Giovannini:2001hh} and contains
automatically the massless U(1) gauge field, which is  a mixture of
the the graviton  fluctuation $h_{\mu\theta}$ and the original U(1)
gauge field fluctuation $A_\mu$ field forming the Nielsen-Olesen
vortex.  This mode was found by direct computation in 
\cite{Giovannini:2002sb,Giovannini:2002mk} and by symmetry arguments
in \cite{Randjbar-Daemi:2002pq}. 

It is known that fermion interaction with topological defects leads
to the existence of localised fermion zero modes
\cite{Jackiw:1975fn,Jackiw:1981ee}. Thus, if fermions are added to
the gravitating Abelian Higgs model in six dimensions, the low energy
effective field theory is expected to be the four-dimensional quantum
electrodynamics with 4d gravity.

The aim of the present paper is to construct explicitly such an
effective theory. Fermionic zero modes in the absence of gravity and
in 4 dimensions in vortex background have been studied by Jackiw and
Rossi in~\cite{Jackiw:1981ee}. Extending this work to six dimensions
is not entirely straightforward. For one thing, the zero mode of
ref.~\cite{Jackiw:1981ee} is localised with the help of a
Majorana-type Yukawa interaction, the structure of which is somewhat
different in $D=6$. Secondly, unlike the flat space case, the
presence of gravity, as we will see, allows the existence of fermion
zero modes even in the absence of Yukawa interactions.  Furthermore,
the Nielsen-Olesen string without gravity does not contain any
localised gauge field, so that the effective field theory contains
nothing but free massless fermions \cite{Hughes:fa}. 

We shall start from an Abelian Higgs model with fermions in $D=6$ and
obtain quantum electrodynamics in 4-dimensional space-time, with
fermionic and gauge wave-functions spread in transverse direction in
a small region in the vicinity of the core of the vortex.  We shall
show that with a judicious choice of variables our zero-mode fermion
equations can be reduced to a form similar, but not identical, to
those of~\cite{Jackiw:1981ee}. We will see that a vector-like theory
in six dimensions leads to a vector-like QED in four dimensions. At
the same time, if one starts from a genuine chiral, but anomaly free
theory in $D=6$, the resulting $D=4$ QED of zero modes in general
contains chiral anomalies. In other words, the six-dimensional theory
can be considered as a self-consistent regularisation of an anomalous
U(1) gauge theory in four dimensions. We also study  the bulk gauge
and fermion spectrum and show that fluctuations of the gauge field
have no mass gap, but those of the charged fermion modes are
massive. The existence of fermion mass gap makes it more plausible to
consider the resulting low-energy theory as a consistent
four-dimensional electrodynamics. 

The paper is organized as follows. In the second section we describe
the bosonic sector of the model: the background solution and the
localised gauge field. In the third section we construct fermion zero
modes for a model in which there is an interaction between fermions
and a scalar. In the fourth section we find interaction between
fermions and the localised gauge field. In section 5 we will consider
localisation of fermions by gravity in the absence of Yukawa coupling
and discuss anomalies. In Section 6 we discuss the bulk gauge and
fermion fields and show the absence of mass gap for vectors and its
presence in the charged fermion spectrum. Section 7 contains our 
conclusions.

\section{The model}
\subsection{The action}
We start from a gravity-Maxwell system in $D=6$, coupled to a complex
scalar field $\Phi$ of charge $e$ and two chiral fermions 
\begin{equation}
\Psi_1= \frac{(1+\Gamma_7)}{2}\Psi_1,~~~~~
\Psi_2=\frac{(1-\Gamma_7)}{2}\Psi_2
\end{equation}
in $4_+$ and $4_-$ representations of SO(1, 5) with U(1)-charges
$e_1$ and $e_2$. Here $\Gamma_7$ is the chirality matrix in $D=6$.
The action is
\begin{eqnarray}
S &=&\int{\rm d}^6 x\sqrt{-G}\left\{{1\over\kappa^2}R
-{1\over 4}F_{MN}F^{MN}-(D_M\Phi)^\dagger D^M\Phi-U(\Phi)
\right.\nonumber\\
&&\qquad
\left.+\sum_{i=1}^2\bar\Psi_i\Gamma^AE_A^M  \nabla_M\Psi_i+
g\bar\Psi_1\Phi\Psi_2+
\mbox{h.c.}\right\}~,
\label{eq1}
\end{eqnarray}
where $D_M\Phi=\partial_M\Phi+i e A_M\Phi$ and
$\nabla_M\Psi_i=(\partial_M-\Omega_M+i e_iA_M)\Psi_i$. Here
$\Omega_M={1\over 2}\Omega_{M[AB]}\Sigma^{AB}$ is the spin connection
which takes its value in the Lie algebra of SO(1, 5) with  generators
$\Sigma_{AB}={1\over 4}[\Gamma_A,\Gamma_B]$  which along with 
$\Gamma_A$ are six $8\times 8$ curved space Dirac matrices with
anticommutational relation  $\{\Gamma_A,\Gamma_B\}_+=2\eta_{AB}$,
$\Gamma_7 = diag(1,-1)$. The indices $M,N$ run from $0$ to $5$,
indices $\mu,\nu$ correspond to 4-dimensional space, and the
signature of the metric is chosen to be $(-,+,\dots,+)$. The model is
free from gravitational, gauge and mixed anomalies only for
$e_1^2=e_2^2$. We choose $e_1=-e_2$, the other option $e_1=e_2$ is
the same as the first one since $4_+$ representation of $\mbox{SO(1,
5)}$ is equivalent to $4_+^*$. The Yukawa-type coupling in
(\ref{eq1}), which can be taken to be real, is non-zero only for
$e_1-e_2=e$. So, the charge assignment we take in Sections 3 and 4 of
the paper is $e_1=e/2,~e_2=-e/2$. The general case of multiple
fermionic species and arbitrary fermionic charges consistent with
anomaly cancellation in $D=6$ will be considered in Section 5. 

In fact, a gauge-invariant mass term
\begin{equation}
L_M=M \overline{\Psi_1^c}\Psi_2 + \mbox{h.c.}~,
\label{majorana}
\end{equation}
where $\Psi^c = C\bar{\Psi}^T$, and $C$ is the matrix of charge
conjugation, can be added to the action (\ref{eq1}). We shall not
include it for the analysis of the zero mode structure, but will
comment on its influence on effective field theory at the end of the
Section 3. This term breaks the fermionic number conservation and can
be forbidden if the fermion number conservation is imposed.

The chiral spinors $\Psi_1$ and $\Psi_2$ can be unified in a single
eight-component spinor as $\Psi = \Psi_1+\Psi_2$, which does not have a
well defined $U(1)$ charge unless $e_1=e_2$,  and the fermionic part of
the Lagrangian (\ref{eq1}) can be written in another form,
\begin{equation}
L_F= \bar\Psi\Gamma^A E_A^M (\partial_M-\Omega_M+ i e_f\Gamma_7 A_M)\Psi+
g\bar\Psi\frac{(1-\Gamma_7)}{2}\Psi\Phi + \mbox{h.c.}~.
\end{equation}
The mass term (\ref{majorana}) in these notations is simply
\begin{equation}
L_M=M \overline{\Psi^c}\frac{(1-\Gamma_7)}{2}\Psi + \mbox{h.c.}~.
\label{maj}
\end{equation}
To get another form of this Lagrangian, which makes its
vector-like structure and thereby the absence of anomalies  evident,
one can introduce a genuine Dirac spinor in six dimensions $\Psi_D =
\Psi_1+\Psi_2^c$. In these notations, the fermionic part of the
Lagrangian is:
\begin{equation}
L_F= \bar\Psi_D\Gamma^A E_A^M (\partial_M-\Omega_M+ i e_f A_M)\Psi_D+
g\bar\Psi_D^c\Gamma_7\Psi_D\Phi + M\bar{\Psi}_D\Psi_D + \mbox{h.c.}~.
\end{equation}
In the paper we are going to use the form (\ref{eq1}-\ref{maj}), as it
allows a simpler treatment of fermionic zero modes.

\subsection{Vortex solution}
It has been shown in~\cite{Giovannini:2001hh} that the bosonic
equations derived from (\ref{eq1}) admit a Nielsen-Olesen vortex-type
solution for which the various field configurations are
\begin{eqnarray}
{\rm d}s^2&=&{\rm e}^{A(r)}\eta_{\mu\nu}{\rm d}x^\mu{\rm d}x^\nu
+{\rm d}r^2+{\rm e}^{B(r)}a^2{\rm d}\theta^2~,
\label{eq2}\\ 
\nonumber
\Phi &=&
f(r){\rm e}^{{\rm i}n\theta},\qquad aeA_\theta\;=\;(P(r)-n){\rm
d}\theta~,
\end{eqnarray}
where $\eta_{\mu\nu}$ is the flat metric, $a$ is the radius of the
circle covered by the coordinate $\theta\in[0,2\pi)$. The integer $n$
is the vortex number and, as in the flat space, the functions $f(r)$
and $P(r)$ satisfy the boundary conditions
\begin{eqnarray}
f(0) =0~, && f(\infty)=f_0\neq 0~,\\
\nonumber
P(0)=n~, &&P(\infty )= 0.
\end{eqnarray}
As $r\to\infty$, $\Phi$ approaches a minimum of the potential
$U(\Phi)$ in (\ref{eq1}). The boundary conditions on the metrical
functions $A(r)$ and $B(r)$ introduced in (\ref{eq2}) are
\begin{eqnarray}
&&A(0)=1,\qquad B(r\to 0)=2\ln\frac{r}{a}~,\\
\nonumber
&&A(r \to \infty)=B(r \to \infty)=-2cr,\quad c>0.
\label{c}
\end{eqnarray}
The parameters $c$ and $a$ are the combinations of the Newton
constant $\kappa$  in the bulk,  the $D=6$ cosmological constant
(related to the value of the scalar potential $U(\Phi)$ at the
minimum), and of the parameters of the Abelian Higgs model
\cite{Gherghetta:2000qi,Giovannini:2001hh}. 

As $r\to 0$, we recover the flat space geometry at the core of the
vortex. Away from the core, the geometry is curved. In particular at
$r \to\infty$, the metric does not  become flat Minkowski and is in
fact the ADS space. It has been shown
in~\cite{Gherghetta:2000qi,Giovannini:2001hh} that this configuration
localises the gravitational fluctuations to the 4-dimensional
subspace spanned by $x^\mu$ at the core of the vortex.

\subsection{Localisation of gauge fields}
In~\cite{Randjbar-Daemi:2002pq} we made a detailed analysis of the
spectrum of fields of spins 0, 1 and 2 in a warped geometry and on
non-trivial gauge and scalar field backgrounds. In particular, we
have given a specific mixture of the fluctuation of the vector
potential and the $\theta\mu$ component of the metric which is
massless in  $D=4$  and has a normalisable action. This
configuration, derived from a symmetry argument similar to the one
given in~\cite{Randjbar-Daemi:1982hi} for the SU(2) case, is given by
\begin{equation}
V_\mu={1\over ae}P(r)W_\mu(x, r),\quad h_{\mu\theta}={\rm
e}^{B(r)}W_\mu(x,r),
\label{eq4}
\end{equation}
where $W_\mu$ is a function of $x^{\mu}$ and $r$.  Substitution of
(\ref{eq4}) in the spin-1 part of the quadratic action, given by
equations (38) to (41) of~ \cite{Randjbar-Daemi:2002pq}, gives rise to an
effective action for the vector $W_\mu$ field:
\begin{equation}
S(W)= -\frac{1}{2 a^2}{2\pi a\over e^2}\int_0^\infty{\rm d}r\,{\rm
e}^{{1\over 2}B} \left(P^2(r)+{a^2e^2\over \kappa^2}{\rm e}^B\right)~
\int{\rm d}^4x[(\partial_\mu  W_\nu)^2 + e^{-A} \partial_r
W_\mu\partial_r W_\mu].
\label{eq3'}
\end{equation}

The $r$ independent $W_\mu$ corresponds to localised massless vector
fields. Their effective action is 
\begin{equation}
S(W)= -\frac{1}{2 q^2} \int{\rm d}^4x(\partial_\mu 
W_\nu)^2.
\label{eq3}
\end{equation}
 where the four-dimensional gauge coupling is
\begin{equation}
\frac{1}{q^2}=
{2\pi a\over e^2}\int_0^\infty{\rm d}r\,{\rm e}^{{1\over 2}B}
\left(P^2(r)+{a^2e^2\over \kappa^2}{\rm e}^B\right)~.
\label{coupl}
\end{equation}
With our boundary conditions on $B(r)$ and $P(r)$, the integral over
$r$ converges at both ends.  A physical interpretation of
(\ref{eq3},\ref{coupl}) is that the fluctuation energy of the
$\mu$-component of the 6-dimensional Maxwell field and the
Kaluza-Klein vector field  $h_{\mu\theta}$ are localised to a region
$r \sim max(1/c,1/M_W,1/M_H)$ near the core of the vortex, where
$M_W$ and $M_H$ are the bulk vector and scalar masses in the
absence of gravity and $c$ is the parameter of the solution, defined
in (\ref{c}). The effective localised field can be taken to be
\[
 \left(P^2(r)+{a^2e^2\over n^2}{\rm e}^B\right)^{1\over 2}
W_\mu(x),
\]
which approaches $W_\mu(x)$ as $r\to 0$ and vanishes rapidly as
$r\to\infty$. It is important to note that the gravitational field
plays a key role in ensuring the normalisability of the effective
action in (\ref{eq4}), through the factor of 
${\rm e}^{{1\over 2}B}$. 
 
\section{Fermions} 
\subsection{Dirac equation in string background} 
As a first step in construction of effective field theory for
fermions one has to find fermionic zero modes in the string
background. This problem has been solved for a Nielsen-Olesen string in
4-dimensions in the absence of gravity in \cite{Jackiw:1981ee}. In
six dimensions and without gravity similar consideration for
supersymmetric Abelian Higgs model has been carried out in
\cite{Hughes:fa}; in \cite{Libanov:2000uf,Frere:2000dc} fermionic
modes were considered for a global string, without gravity and gauge
fields. In \cite{Neronov:2001qv} gravity has been taken into account,
but interaction of fermions with gauge and Higgs was not included. In
this section we will find fermionic zero modes on a six-dimensional
string incorporating gravity as well as gauge-Higgs interactions, a
special cases of the general problem which was addressed in
\cite{Randjbar-Daemi:2000cr}.

General equations for the fermions in the warped metric
and in the presence of gauge and Higgs fields can be found, e.g. in
\cite{Randjbar-Daemi:2000cr}. Applied to our case, the Dirac equation
reads:
\begin{equation}
\left\{
e^{-A/2}\Gamma^{\underline \mu}  \partial_\mu + \Gamma^{\underline r}
\left(\partial_r + A' +\frac{B'}{4}\right)
+e^{-\frac{B}{2}}\Gamma^{\underline\theta}
\left(\frac{1}{a}\partial_\theta
+i\frac{e}{2}\Gamma_7A_\theta\right)+\right.
\label{dirac}
\end{equation}
\[
\left. g\frac{(1-\Gamma_7)}{2}\Phi +g\frac{(1+\Gamma_7)}{2}\Phi^*
\right\}\Psi =0~,
\]
where prime denotes the derivative with respect to $r$, and underlined
indices correspond to flat space coordinates. Here we take $g \neq 0$
and $e_1=-e_2=e/2$.

A special care should be taken on the boundary conditions of
fermionic wave function with respect to $\theta$.  Because of the
veilbein transformation properties it has to be {\it antiperiodic},
$\Psi(\theta)=-\Psi(\theta+2\pi)$. This corresponds to a single valued
spinor in Cartesian coordinate system near $r\rightarrow 0$.

To proceed, we fix the $D=6$ Dirac matrices in the following form:
$\Gamma_{\underline \mu}=\gamma_\mu\times\sigma_1$,
$\Gamma_{\underline{r}}=\gamma_5\times \sigma_1$ and 
$\Gamma_{\underline{\theta}}=1\times \sigma_2$, where $\sigma_1$ and
$\sigma_2$ are Pauli matrices and $\gamma_\mu$ are $4\times 4$ Dirac
matrices chosen in such a way that $\gamma_5$ is diagonal, 
i.e.~$\gamma_5=diag(1,-1)$. Taking into account that 
\begin{equation}
\Gamma_{\underline{r}\underline{\theta}} \equiv \Gamma_{\underline r}
\Gamma_{\underline\theta} = \left(\begin{array}{cc} i\gamma_5 &
0\\[2mm] 0 & -i\gamma_5 \end{array}\right)~,
\end{equation}
it is convenient to introduce a new  field $\psi$ through the definition,
\begin{equation}
\Psi = 
\exp\left\{\frac{\theta}{2}    
\Gamma_{\underline{r}\underline{\theta}} -A-{1\over 4}B +
\int^r{\rm d}\rho{\rm e}^{-B(\rho)/2}
\left({1\over 2a}-{\rm i}\Gamma_7{e\over 2}
\Gamma_{\underline{r}\underline{\theta}}
A_\theta(\rho)\right)\right\}\psi~,
\label{change}
\end{equation}
where
\begin{equation}
\psi=\left(\begin{array}{c}
\psi_1 \\ \psi_2 \end{array}\right).
\end{equation}
Note that due to the prefactor $\exp
{(\frac{\theta}{2}\Gamma_{r\theta})}$ and the antiperiodicity of
$\Psi$ the new field $\psi$ must be  {\it periodic}  in $\theta$.
  
Substitution of eq.(\ref{change}) to (\ref{dirac}) removes the gauge
field from the equations and yields the equations for 4d zero modes
$\gamma^{\mu}  \partial_\mu \psi_i=0$:
\begin{eqnarray}
{\rm e}^{+{\rm i}\theta\gamma_5}\left(\partial_r+{{\rm i}\over a}
\gamma_5{\rm
e}^{-B/2}\partial_\theta\right)\psi_1+g\Phi\gamma_5\psi_2 &=&0,\\
\nonumber
{\rm e}^{-{\rm i}\theta\gamma_5}\left(\partial_r-{{\rm i}\over a}
\gamma_5{\rm
e}^{-B/2}\partial_\theta\right)\psi_2+g\Phi^*\gamma_5\psi_1 &=&0.
\end{eqnarray}

To understand better their chiral structure, we write $\psi_i$ in
terms of $D=4$ left and right handed spinors, 
\begin{equation}
\psi_i = \left(\begin{array}{c}\psi_i^L\\ \psi_i^R
\end{array}\right)~,
\label{4dlr}
\end{equation}
and get the system of four equations for the four different chiral
components. For the left spinors we have
\begin{eqnarray}
{\rm e}^{+{\rm i}\theta} \left(\partial_r+{{\rm i}\over a}
{\rm e}^{-B/2}\partial_\theta\right)\psi_1^L+g\Phi~\psi_2^L&=&0,\\
\nonumber
{\rm e}^{-{\rm i}\theta} \left(\partial_r-{{\rm i}\over a}
{\rm e}^{-B/2}\partial_\theta\right)\psi_2^L+g\Phi^*\psi_1^L&=&0,
\end{eqnarray}
and, for the right spinors correspondingly
\begin{eqnarray}
{\rm e}^{-{\rm i}\theta} \left(\partial_r-{{\rm i}\over a}
{\rm e}^{-B/2}\partial_\theta\right)\psi_1^R-g\Phi~\psi_2^R&=&0,\\
\nonumber
{\rm e}^{+{\rm i}\theta} \left(\partial_r+{{\rm i}\over a}
{\rm e}^{-B/2}\partial_\theta\right)\psi_2^R-g\Phi^*\psi_1^R&=&0~.
\end{eqnarray}

These equations are similar but not identical to those of
\cite{Jackiw:1981ee}.
\subsection{Localised fermion zero modes}
The four-dimensional effective Lagrangian for fermion reads 
\begin{eqnarray}
\nonumber
\int dr d\theta \sqrt{-G}
\sum_{i=1}^2\bar\Psi_i\Gamma^AE_A^M  \nabla_M\Psi_i=
\int dr d\theta 
\sum_{i=1}^2\left\{N_L(r)\bar\psi_i^L\gamma^\mu\partial_\mu\psi_i^L+
N_R(r) \bar\psi_i^R\gamma^\mu\partial_\mu\psi_i^R
\right\}~,
\label{efflagr}
\end{eqnarray}
where
\begin{eqnarray}
N_L(r)=\exp\left[-\frac{A}{2}+ \int^r{\rm d}\rho{\rm e}^{-B(\rho)/2}
\left({1\over a}+e
A_\theta(\rho)\right)\right],
\label{normg}
\\
\nonumber
N_R(r)=\exp\left[-\frac{A}{2}+ \int^r{\rm d}\rho{\rm e}^{-B(\rho)/2}
\left({1\over a}-e
A_\theta(\rho)\right)\right].
\end{eqnarray}
This fixes the condition of normalisation of fermion zero modes. The
behaviour of functions $N_{L,R}$ at zero and infinity are as follows:
\begin{eqnarray}
\nonumber
N_{L,R}(r \to 0) &\to& r~,\\
N_L(r \to \infty) &\to& \exp\left(cr+\frac{1-n}{ac}e^{cr}\right)~,
\label{norm}
\\
\nonumber
N_R(r \to \infty) &\to& \exp\left(cr+\frac{1+n}{ac}e^{cr}\right).
\end{eqnarray}
In the following, we take the string winding number $n$ to be positive.

The $\theta$ dependence can be removed by the substitutions
\begin{equation}
\psi_1^L=e^{im\theta}u_m^L(r)\chi_m^L(x^\mu),~
\psi_2^L=e^{i(m+1-n)\theta}v_m^L(r)\chi_m^L(x^\mu),
\label{leftsub}
\end{equation}
and
\begin{equation}
\psi_2^R=~e^{im\theta}u_m^R(r)\chi_m^R(x^\mu),~
\psi_1^R=-e^{i(m+1+n)\theta}v_m^R(r)\chi_m^R(x^\mu)~,
\end{equation}
where $\chi_m^{L,R}(x^\mu)$ are two-component $SO(1,3)$
spinors. 
The radial wave functions can be found from
\begin{eqnarray}
\left(\partial_r -\frac{m}{a}e^{-B/2}\right)u_m^L +gf v_m^L=0~,
\label{jr}\\
\nonumber
\left(\partial_r -\frac{n-1-m}{a}e^{-B/2}\right)v_m^L +gf
u_m^L=0~.
\end{eqnarray}
for the $D=4$ left fermions and from
\begin{eqnarray}
\left(\partial_r -\frac{m}{a}e^{-B/2}\right)u_m^R +gf v_m^R=0~,
\label{jl}\\
\nonumber
\left(\partial_r +\frac{n+1+m}{a}e^{-B/2}\right)v_m^R +gf
u_m^R=0~.
\end{eqnarray}
for the $D=4$ right fermions. These equations reduce exactly to the
ones of  refs. \cite{Jackiw:1981ee,Libanov:2000uf}  for the flat
case, when $B = 2 log(r/a)$. They have the symmetry $m
\leftrightarrow n-1-m,~u \leftrightarrow v$ for the left fermions and
the symmetry $m\leftrightarrow -n-1-m,~u \leftrightarrow v$ for the
right ones. This property will be used later to combine chiral fields
to $D=4$ Dirac fermions.

The behaviour of the modes at small $r$ follows immediately from
\cite{Jackiw:1981ee}: solutions for left-handed fermions are always
regular provided $m$ takes integer values $m=0,\dots,n-1$, whereas
one of the independent solutions for right-handed modes is always
singular at the origin. So, we concentrate on large $r$ limit, where
$B\to 2cr$ and $gf \to M_f$, where $M_f$ is the fermion mass in the
bulk. 

Consider first left-handed modes. Introducing $x=\frac{e^{cr}}{ac}$,
$m_f=M_f/c>0$, and $u_{L}=x^{m_f}e^{mx}y(x)$, equation for $y(x)$ is
the one for a degenerate hypergeometric function,
\begin{equation}
xy'' +\left(1+2m_f -(n-2m-1)x\right)y' -m_f(n-2m-1)y=0~.
\label{hyper}
\end{equation}
If $m=\frac{n-1}{2}$ which is possible for odd $n$, the solution is
simply
\begin{equation}
y=C_1+C_2x^{-2m_f}~.
\end{equation}
The normalisability condition gives $C_1=0$ and requires $m_f \neq
0$. This solution corresponds to a left-handed chiral fermion
localised on a string. If $m_f = 0$, the mode is not normalisable,
since the normalisation integral diverges as $exp(cr)$\footnote{In
examining the normalisability of a solution we need to include the
contribution of the $r$- dependent prefactor in eq.
(\ref{normg},\ref{norm}).}.

For $m \neq\frac{n-1}{2}$ the $x\rightarrow\infty$ asymptotics of the 
general solution to (\ref{hyper}) is:
\begin{equation}
y=C_1 x^{-m_f}+C_2 e^{(n-1-2m)x}x^{-m_f-1}~,
\end{equation}
what gives for $u_{L}$
\begin{equation}
u_{L} \to C_1 e^{mx} + C_2 e^{(n-1-m)x}x^{-1}~.
\label{ulsol}
\end{equation}
From here and (\ref{norm}) it is clear that the left-handed fermionic
modes with $m=0,\dots,n-1$ are always normalisable: to get a
normalisable left mode one should take $C_2$=0 for $m<\frac{n-1}{2}$
and  $C_1$=0 for $m >\frac{n-1}{2}$. 

For the analysis of the right-handed modes it is sufficient to change
$n$ to $-n$ in eqs. (\ref{hyper}-\ref{ulsol}). These modes are not
normalisable at any choice of $m$. 

Finally, the effective Lagrangian is
\begin{equation}
L= \sum_{m=0}^{n-1} N_m\bar\chi_m^L\gamma^\mu\partial_\mu\chi_m^L~,
\label{ll}
\end{equation}
where
\begin{equation}
N_m = 2 \pi \int dr N_L(r)\left(|u_m|^2+|v_m|^2\right)~.
\label{nm}
\end{equation}

To summarise: for $g\neq 0$ we have exactly $n$ left-handed
normalisable fermionic zero modes.  So far we have assumed that n is
positive. If $n$ is negative we get the same pattern for the
localised right-handed modes.

The inclusion of the bulk Majorana-type mass (\ref{majorana}) adds the
following term to the four-dimensional effective Lagrangian:
\begin{equation}
M {\overline{\psi_{1L}^c}}\psi_{2L} + \mbox{h.c.}~.
\end{equation}
In this case, the modes with number $m$ can be put together with 
charge-conjugated modes $(n-1-m)$ to form a massive Dirac spinor,
whereas the mode $m=\frac{n-1}{2}$ is a Majorana fermion (for $n$
odd). As we will see in Section 6, the spectrum of {\it charged} bulk
fermions develops a mass gap, and, therefore, four-dimensional 
massive charged spinors are the genuine localised states. On the
contrary, for neutral Majorana fermions the mass gap is absent (see
Section 6 and also \cite{Dubovsky:2000am}), and it represents in fact
a metastable state.

\section{ U(1) charges in $D=4$}
Our aim in this section is to define the interaction between the
localised fermionic modes and the gauge field $W_\mu(x)$, defined in
(\ref{eq4}). While the interaction of a part of $W_\mu$ coming from
the initial gauge field $V_\mu$ can be written down immediately, the
gravity part, related to the component of the metric $h_{\mu\theta}$
requires some care. To find the coupling of the ``graviphoton" to
$\psi$, one can  calculate the spin connections $\Omega_M$ starting
from the metric
\begin{equation}
ds^2 = e^{A(r)} \eta_{\mu\nu} dx^\mu dx^\nu +dr^2 + e^{B(r)}
\left(ad\theta+W_{\mu}(x) dx^\mu\right)^2~.
\label{me}
\end{equation}
Using standard formulae we can calculate the components of
$\Omega_A$.  They are given more compactly in terms of the frame
components $\Omega_A =E_A\ ^M \Omega_M$.  We choose the frames
\begin{equation}
 E_a\ ^\mu = e^{-A/2} \delta_a\ ^\mu~,\qquad E^\theta_\mu =
-e^{-A/2}W_\mu~,
\label{conn}
\end{equation}
\[
 E^r_{\underline r} = 1~,\qquad E^\theta_{\underline\theta} =
e^{-B/2}~,
\]
where $a=0,1,2,3$ and an underlined character refers to the
orthonormal frame.  The non-vanishing components are
\begin{equation}
\Omega_a = {1\over 4} A' \Gamma_{{\underline r} a} -
{1\over 4} e^{B/2} W_{ab}
\Gamma^{b\underline\theta}~,
\end{equation}
\[
\nonumber
\Omega_{\underline\theta} = {1\over 4} B' \Gamma_{{\underline r} 
\underline\theta}
-{1\over 4} e^{B/2} W_{ab}
\Sigma^{ab}~,
\]
where $W_{ab}=e^{-A}\delta_a\ ^\mu \delta_b\ ^\nu$ ($\partial_\mu
W_\nu -\partial_\nu W_\mu)$. The $W_{ab}$ containing terms could give
rise to the tree level magnetic moment couplings in 4--dimensions. 
In our model, due to the $D=6$ chirality, they drop out.

Upon substitution from (\ref{change}), (\ref{me}) and (\ref{conn}) in
(\ref{eq1}) we obtain the action,
\begin{equation}
S_F = \int d^6 x\left[ N_{\varepsilon_1}(r) \bar\psi_1\gamma^\mu
(\partial_\mu + \frac{i}{a}W_\mu Q_1)\psi_1 + N_{\varepsilon_2}(r)
\bar\psi_2\gamma^\mu (\partial_\mu + \frac{i}{a}W_\mu
Q_2)\psi_2\right]~,
\label{SF}
\end{equation}
where $\varepsilon_{i}^2 =1$ define the chirality, i.e.
$\gamma_5\psi_i=\varepsilon_i\psi_i$ with  $N_{+1}=N_L$, 
$N_{-1}=N_R$, and
$N_{L,R}$ are defined in (\ref{normg}). The $U(1)$ charge operator
acting on $\psi_1$ is defined by
\begin{equation}
Q_1 = i\partial_\theta -\frac{\varepsilon_1}{2}+ \frac{n}{2}
\label{Q1}
\end{equation} 
and the one acting on $\psi_2$ is given by
 \begin{equation}
Q_2 = i\partial_\theta +\frac{\varepsilon_2}{2}+ \frac{n}{2}~.
\label{Q2}
\end{equation} 

Written in terms of $\chi_L$ introduced in (\ref{leftsub})  the
trilinear part of the effective Lagrangian contain $\chi_L$ and
$W_\mu$ becomes 
\begin{equation}
L_{int}= \frac{i}{a}\sum_{m=0}^{n-1} 
Q_m\bar\chi_m^L\gamma^\mu \chi_m^L W_\mu~,
\end{equation}
where 
\begin{equation}
Q_m = 2 \pi \left(m-\frac{n-1}{2}\right)\int dr N_L(r)
\left(|u_m|^2+ |v_m|^2 \right)~.
\end{equation}
It is essential, that the part of the localised U(1) gauge field,
coming from $V_\mu$ and proportional to $P(r)$, cancels out. Using
(\ref{ll}), (\ref{nm}), (\ref{SF}) and (\ref{Q1}) 
the $D=4$ effective Lagrangian becomes
\begin{equation}
L= \sum_{m=0} ^{m=n-1} N_m \bar\chi_m \gamma^\mu (\partial_\mu +
\frac{i}{a} q_m W_\mu)\chi_m
\end{equation}
 Therefore, the four-dimensional charges of fermions are simply
\begin{equation}
q_m= \left(m-\frac{n-1}{2}\right)q~,
\end{equation}
where $q$ is defined in (\ref{coupl}). This means that the chiral
fermion with $m=\frac{n-1}{2}$ (for odd $n$) is neutral. We also
notice that $\chi_m$ can be unified with $\chi_{n-1-m}^c$ to form a
Dirac spinor.

Thus, for the even vortex number $n=2k$, we find $k$ Dirac fermions
with charges $\frac{1}{2},\dots, k-\frac{1}{2}$, whereas for an odd
vortex number $n=2k+1$ one gets $k$ charged Dirac fermions and one
neutral Weyl fermion. Such a  Weyl fermion would give rise to a
gravitational anomaly in two, six or ten dimensions, but has an
anomaly free coupling to gravity in a four-dimensional space-time. As
a result, a vector-like theory in $D=6$ leads to a vector-like QED in
$D=4$. It is interesting to note that for even $n$ the charges are
quantised in units of half-odd integers, while for odd $n$ we obtain
integrally quantised charges.
\section{Gravitational localisation of fermions\\ and anomalies}
If there is no interaction between fermions and the scalar,  the
values of $D=6$ left and right handed  fermionic charges can be
arbitrary, up to the requirement of anomaly cancellation in six
dimensions. If we have $n_L$ left fermions with charges $e_L^i$ and
$n_R$ right-handed fermions with charges $e_R^i$, the absence of
gravitational anomalies requires $n_L =n_R \equiv n_F$, the absence
of gauge anomalies gives $\sum (e_L^i)^4= \sum (e_R^i)^4$, and the
absence of mixed anomalies leads to $\sum (e_L^i)^2= \sum (e_R^i)^2$
\cite{Alvarez-Gaume:1983ig}. For $n_F=1,2$ these conditions lead
necessarily to a vector-like theory in $D=6$, whereas for $n_F\geq 3$
a genuine chiral gauge theory is allowed\footnote{We are not assuming
that the values of the e's are necessarily quantized.}.

To analyse the general case, it is sufficient to put the Yukawa
coupling  $g=0$ in all equations of the previous sections and remove
the  constraint relating the charges of the fermions and  the scalar.
Without loss of generality all charges can be assumed to be positive,
as the fermion with negative charge can be transformed into a fermion
with the positive charge by operation of charge conjugation which,
due to the pseudo complexity of the chiral spinor representation of
$SO(1,5)$, commutes with chirality in $D=6$. We take, as usual, the
winding number $n>0$.  Since the results are obvious from trivial
modifications of  eqs. (\ref{efflagr}-\ref{jr}), we omit the
details. 

For each left-handed (right-handed) six-dimensional spinor with
charge $e_L^i$ ($e_R^i$) one has  {\it left-handed} four-dimensional
Weyl spinors with normalisable (accounting for the factor
(\ref{norm})) wave-functions
\begin{equation}
\psi=\exp(\pm im\theta)\exp\left[\frac{m}{a} \int^r{\rm
d}\rho{\rm e}^{-B(\rho)/2}\right]\chi(x)~,
\end{equation}
where, m in the exponent stands for $m_L$ or $m_R$  which assume  
integer values in the range
$ m_R= 0, 1, 2 ... <
 \frac{e_R^i}{e}n-\frac{1}{2}$ and $ m_L= 0, 1, 2 ... <
 \frac{e_L^i}{e}n-\frac{1}{2}$. Also  
the plus (minus) sign  in the exponent refers to  the left (right)
$D=6$ spinor chirality. The $D=4$ electric charges of localised
spinors can be read by acting the charge operators $Q_1$ or $Q_2$
given above in (\ref{Q1},\ref{Q2}) on $\psi$. The result is:
\begin{equation}
\frac{1}{2} + m_L -n\frac{e_L^i}{e}~~ \mbox{and}~~ -\frac{1}{2} - m_R
+n\frac{e_R^i}{e}~.
\end{equation}
The region where $D=4$ fermions are localised is related to $c$ and
is of the order of $r \sim \frac{1}{c}$.

For $n_F=1,2$ the effective 4D theory is necessary vectorlike, but
for $n_F\geq 3$ the resulting theory can be chiral and anomalous. As
an example, consider the string with winding number $n=1$ and the
following  anomaly-free in $D=6$ charge assignment:
$e_1^L=0,~e_2^L=1, e_3^L=1$ and  $e_1^R=2/\sqrt{3},~e_2^R=1/\sqrt{3},
e_3^R=1/\sqrt{3}$, $e=1$. Normalisable solutions do not exist for the
$D=6$ fermion with $e_1^L=0$, while for other fermions only the
choice of  $m_L=m_R=0$ leads to localised $D=4$ fermions. So, in
$D=4$ we have five left-handed fermions with charges
$-1/2,~-1/2,~(2/\sqrt{3}-1/2),~(1/\sqrt{3}-1/2)$ and
$~(1/\sqrt{3}-1/2)$. The four-dimensional anomaly cancellation
equation $\sum (q_i^L)^3=0$ is not satisfied. The fact that anomalous
U(1) gauge theory in four dimensions can be constructed as an
effective theory out of an anomaly free theory in higher dimensions
means that an abelian anomalous gauge theory can be made
mathematically consistent. Indeed, there has been some attempts in
the past (see, e.g. \cite{Faddeev:pc} and references therein) to make
physical sense out of anomalous chiral gauge theories.  In our scheme
the $D=4$ theory is a low energy approximation to a bigger and
consistent theory in $D=6$. In this sense our construction can be
considered as a regularization of $D=4$ anomalous chiral gauge
theories. The restoration of $D=4$ gauge invariance should come from
the fact that the  excitations of heavy fermions living in the bulk
must be essential and cannot be discarded, in analogy with ref.
\cite{D'Hoker:1984pc}. However, the study of the question of how this
happens exactly, goes beyond the scope of this paper. 

\section{The mass gaps} 

As we have already discussed, a theory of fermionic and gauge zero
modes, localised on a string, can be a ``good" four-dimensional
effective field theory provided bulk modes interact weakly with the
localised modes at small energies. A complete analysis of this
problem is complicated, and, to our  knowledge has never been
done for any type of brane-world scenario. Our aim in this section is
more modest: we are going to demonstrate that the bulk gauge modes
are not separated from zero modes by a mass gap. At the same time,
the charged bulk fermions are massive, with a mass gap $m_F\sim
\frac{1}{a}$. This means that at small energies, $E < \frac{1}{a}$
our theory is indeed the four-dimensional QED. In particular, the
processes with visible electric charge nonconservation have a
threshhold behaviour, in  contrast with the model of ref.
\cite{Dubovsky:2000av}.
\subsection{Gauge fields}
To define the spectrum of gauge modes one can use general equations 
of \cite{Randjbar-Daemi:2002pq} for spin-1 fluctuations. In our case,
three vector fields are present - two coming from the metric,
$h_{\mu\theta}$ and $h_{\mu r}$, and the third is the U(1) gauge
field making the Nielsen-Olesen string. Our interest is related to the bulk
excitations with photon quantum numbers.  The field, corresponding to
photon, decouples from all other vector fields and has the form of
(\ref{eq4}), where $W_\mu$ is now a function of $r$ and $x^\mu$ (but
not of $\theta$). Since our metric is regular at $r \to 0$, to
understand the structure of the spectrum it is sufficient to consider
the $r\rightarrow \infty$ limit of the corresponding equation for mass
eigenvalues.  In what follows, we will assume that $M_W > c$, so that
$P(r)$ can be neglected in comparison with $e^A$, $e^B$ at large
distances. Then the equation defining the spectrum of bulk photons with
four-dimensional masses $m_w$ follows directly from eq.(42) of 
\cite{Randjbar-Daemi:2002pq} and is, at $r \to \infty$:
\begin{equation}
-e^{-\frac{3}{2}B}\frac{\partial}{\partial r}
\left[e^{\frac{3}{2}B+A}\frac{\partial
W(r)}{\partial r}\right] = m_w^2 W(r)~,
\end{equation}
At large $r$ the asymptotic of the solution is simply a collection of
plane waves
\begin{equation}
W(r) \to z^2 \left(C_1 \sin(z) + C_2 \cos(z)\right)~,
\end{equation}
where $z=\frac{m_w}{c} e^{cr}$, what proves the absence of a mass gap.

\subsection{Fermions}
A similar consideration can be carried out for fermions. For large
$r$, the Yukawa term and Majorana mass in (\ref{dirac}) can be
neglected,  and the equations for bulk modes (\ref{4dlr}) reads (we
consider only $4_+$ six-dimensional fermion, the case of $4_-$ is
treated in full analogy) :
\begin{eqnarray}
{\rm e}^{+{\rm i}\theta} \left(\partial_r+{{\rm i}\over a}
{\rm e}^{-B/2}\partial_\theta\right)\psi_L &=&
- e^{-\frac{1}{2} A -2\int^r d\rho e^{-\frac{B}{2}}e_L A_\theta }
\gamma^\mu\partial_\mu\psi_R~,
\label{mes}
\\
\nonumber
{\rm e}^{-{\rm i}\theta} \left(\partial_r-{{\rm i}\over a}
{\rm e}^{-B/2}\partial_\theta\right)\psi_R &=& 
~~e^{-\frac{1}{2} A +2\int^r d\rho e^{-\frac{B}{2}}e_L
A_\theta}
\gamma^\mu\partial_\mu\psi_L~.
\end{eqnarray}
The ansatz $\psi_L=e^{im\theta}\psi_L^m$,
$\psi_R=e^{i(m+1)\theta}\psi_L^m$ removes the angular dependence and
leads, for large $r$, to the following equation for the spectrum:
\begin{equation}
\left[-\frac{\partial^2}{\partial r^2}
+\frac{2ne_L/e-1}{a}e^{-\frac{B}{2}}
\frac{\partial}{\partial r} +
\frac{m(m+1-2ne_L/e)}{a^2}e^{-B} \right] \psi_L ^m = e^{-A} m_F^2
\psi_L^m~,
\end{equation}
where $n$ is the string winding number and $m_F^2$ are the
eigenvalues of the 4-dimensional mass$^2$ operator $\partial^2$. At
large
$r$, asymptotics of the solutions are
\begin{equation}
\psi_L ^{m} \to \frac{1}{\sqrt{z}} \exp
\left(-\frac{1-2ne_L/e}{2am_F}z\right)  \left(C_1 \sin(\omega z) +
C_2 \cos(\omega z)\right)~,
\end{equation}
where $z=\frac{m_F}{c} e^{cr}$ and
\begin{equation}
\omega^2=1- \frac{(m-ne_L/e+\frac{1}{2})^2}{(am_F)^2} ~.
\label{root}
\end{equation}
The numerator on the right hand side of $\omega^2-1$, is nothing but
the square of the charge operator $Q= i\partial_\theta - \frac{1}{2}+
n\frac{e_L}{e}$ acting on $\psi_L$. It is easy to work out a similar
solution for $\psi_R$ and show that  the states $\psi_R$ with the
same charge as $\psi_L$  can be put  together to construct a
four component massive charged spinor.

From (\ref{mes}) one can see that at small $r$ and $m_F \neq 0$ and
for any value of $m>0$ there is only one regular wave function with
the behaviour $\psi_L ^{m} \simeq 2C(m+1)r^m,~\psi_R ^{m} \simeq -C m_F
r^{m+1}$, where $C$ is an arbitrary constant. This means that the
spectrum of the mass operator is discrete for $m_F <\frac{1}{a}Q$ and
continuous afterwords, see eq. (\ref{root}). Thus, the masses of the 
bulk  Dirac fermions  are bounded by their charge, viz,  $m_F
>\frac{1}{a}Q$, and the charged zero modes are separated from other
modes by a mass gap.

\section{Conclusion and outlook} 
In this paper, starting from an anomaly free $U(1)$ gauge theory in
$D=6$ we have constructed a full fledged effective $D=4$
electrodynamics of charged particles interacting with photons and
gravitons.  Due to gravitational interactions in $D=6$ the theory
dynamically localises massless (or massive, if a Majorana-type mass
term is added in $D=6$) charged fermions and photons to a small
region around the core of a Nielsen-Olsen vortex. 

Of course, there is a very long way from gravitating quantum
electrodynamics to a realistic model incorporating non-Abelian gauge
fields. The U(1) theory we discussed may be considered as a prototype
of such a model. It would be interesting to extend our work to
anomaly free supersymmetric models in $D=6$.  Such models do exist
\cite{Randjbar-Daemi:wc} and they contain Yukawa type couplings
between fermions of different chiralities which played an important
role in our construction.

Another question to ask is what is the low-energy behaviour of an
anomalous $D=4$  chiral gauge theory derived from an anomaly free
theory in $D=6$. In Section 5 we argued that our scheme can be
considered as a regularisation of the anomalous $D=4$ chiral gauge
theories. It is clear that by including a finite number of massive
fermionic modes, which have vector-like couplings, the theory cannot
be made anomaly free. This implies that all the infinite number of
fermionic modes must be incorporated, i.e. for the full consistency
of the theory the heavy fields should not completely decouple from
low energy physics. The study of this question deserves further
investigation.


{\bf Acknowledgments.}  We thank P. Tinyakov and M.
Giovannini for discussions. S.R.D. and M.S. are appreciative of the
hospitality at the IPT of Lausanne and at the ICTP in Trieste
correspondingly. This work was supported by the Swiss Science
Foundation, contract no. 20-64859.01.


\begin{thebibliography}{99}

\bibitem{Rubakov:bb}
V.~A.~Rubakov and M.~E.~Shaposhnikov,
Phys.\ Lett.\ B {\bf 125} (1983) 136.

\bibitem{Akama:jy}
K.~Akama,
Lect.\ Notes Phys.\  {\bf 176} (1982) 267
[arXiv:hep-th/0001113].

\bibitem{Visser:qm}
M.~Visser,
Phys.\ Lett.\ B {\bf 159} (1985) 22
[arXiv:hep-th/9910093].

\bibitem{Dvali:1996xe}
G.~R.~Dvali and M.~A.~Shifman,
Phys.\ Lett.\ B {\bf 396} (1997) 64
[Erratum-ibid.\ B {\bf 407} (1997) 452]
[arXiv:hep-th/9612128].

\bibitem{Arkani-Hamed:1998rs}
N.~Arkani-Hamed, S.~Dimopoulos and G.~R.~Dvali,
Phys.\ Lett.\ B {\bf 429} (1998) 263
[arXiv:hep-ph/9803315].

\bibitem{Randall:1999vf}
L.~Randall and R.~Sundrum,
Phys.\ Rev.\ Lett.\  {\bf 83} (1999) 4690
[arXiv:hep-th/9906064].

\bibitem{Rubakov:2001kp}
V.~A.~Rubakov,
Phys.\ Usp.\  {\bf 44} (2001) 871
[Usp.\ Fiz.\ Nauk {\bf 171} (2001) 913]
[arXiv:hep-ph/0104152].

\bibitem{Polchinski:1995mt}
J.~Polchinski,
Phys.\ Rev.\ Lett.\  {\bf 75} (1995) 4724
[arXiv:hep-th/9510017].

\bibitem{Dubovsky:2001pe}
S.~L.~Dubovsky and V.~A.~Rubakov,
Int.\ J.\ Mod.\ Phys.\ A {\bf 16} (2001) 4331
[arXiv:hep-th/0105243].

\bibitem{Gherghetta:2000qi}
T.~Gherghetta and M.~E.~Shaposhnikov,
Phys.\ Rev.\ Lett.\  {\bf 85} (2000) 240
[arXiv:hep-th/0004014].

\bibitem{Oda:2000zc}
I.~Oda,
Phys.\ Lett.\ B {\bf 496} (2000) 113
[arXiv:hep-th/0006203].

\bibitem{Dubovsky:2000av}
S.~L.~Dubovsky, V.~A.~Rubakov and P.~G.~Tinyakov,
JHEP {\bf 0008} (2000) 041
[arXiv:hep-ph/0007179].

\bibitem{Giovannini:2001hh}
M.~Giovannini, H.~Meyer and M.~E.~Shaposhnikov,
Nucl.\ Phys.\ B {\bf 619} (2001) 615
[arXiv:hep-th/0104118].

\bibitem{Neronov:2001br}
A.~Neronov,
Phys.\ Rev.\ D {\bf 64} (2001) 044018
[arXiv:hep-th/0102210].

\bibitem{Nielsen:qs}
H.~B.~Nielsen and P.~Olesen,
Nucl.\ Phys.\ B {\bf 57} (1973) 367.

\bibitem{Giovannini:2002sb}
M.~Giovannini,
Phys.\ Rev.\ D {\bf 66} (2002) 044016
[arXiv:hep-th/0205139].

\bibitem{Giovannini:2002mk}
M.~Giovannini, J.~V.~Le Be and S.~Riederer,
Class.\ Quant.\ Grav.\  {\bf 19} (2002) 3357
[arXiv:hep-th/0205222].

\bibitem{Randjbar-Daemi:2002pq}
S.~Randjbar-Daemi and M.~Shaposhnikov,
Nucl.\ Phys.\ B {\bf 645} (2002) 188
[arXiv:hep-th/0206016].

\bibitem{Jackiw:1975fn}
R.~Jackiw and C.~Rebbi,
Phys.\ Rev.\ D {\bf 13} (1976) 3398.

\bibitem{Jackiw:1981ee}
R.~Jackiw and P.~Rossi,
Nucl.\ Phys.\ B {\bf 190} (1981) 681.

\bibitem{Randjbar-Daemi:1982hi}
S.~Randjbar-Daemi, A.~Salam and J.~Strathdee,
Nucl.\ Phys.\ B {\bf 214} (1983) 491.

\bibitem{Hughes:fa}
J.~Hughes, J.~Liu and J.~Polchinski,
Phys.\ Lett.\ B {\bf 180} (1986) 370.

\bibitem{Libanov:2000uf}
M.~V.~Libanov and S.~V.~Troitsky,
Nucl.\ Phys.\ B {\bf 599} (2001) 319
[arXiv:hep-ph/0011095].

\bibitem{Frere:2000dc}
J.~M.~Frere, M.~V.~Libanov and S.~V.~Troitsky,
Phys.\ Lett.\ B {\bf 512} (2001) 169
[arXiv:hep-ph/0012306].

\bibitem{Neronov:2001qv}
A.~Neronov,
Phys.\ Rev.\ D {\bf 65} (2002) 044004
[arXiv:gr-qc/0106092].

\bibitem{Randjbar-Daemi:2000cr}
S.~Randjbar-Daemi and M.~E.~Shaposhnikov,
Phys.\ Lett.\ B {\bf 492}, 361 (2000)
[arXiv:hep-th/0008079].

\bibitem{Dubovsky:2000am}
S.~L.~Dubovsky, V.~A.~Rubakov and P.~G.~Tinyakov,
Phys.\ Rev.\ D {\bf 62} (2000) 105011
[arXiv:hep-th/0006046].

\bibitem{Alvarez-Gaume:1983ig}
L.~Alvarez-Gaume and E.~Witten,
Nucl.\ Phys.\ B {\bf 234} (1984) 269.

\bibitem{Faddeev:pc}
L.~D.~Faddeev and S.~L.~Shatashvili,
Phys.\ Lett.\ B {\bf 167} (1986) 225.

\bibitem{D'Hoker:1984pc}
E.~D'Hoker and E.~Farhi,
Nucl.\ Phys.\ B {\bf 248} (1984) 77.

\bibitem{Randjbar-Daemi:wc}
S.~Randjbar-Daemi, A.~Salam, E.~Sezgin and J.~Strathdee,
Phys.\ Lett.\ B {\bf 151}, 351 (1985).

\end{thebibliography}
\end{document}